# Optics-less smart sensors and a possible mechanism of cutaneous vision in nature

Leonid Yaroslavsky[1], Chad Goerzen[1], Stanislav Umansky[1] and H. John Caulfield[2]

[1] *Dept. Physical Electronics, Faculty of Engineering. Tel Aviv University.*
*Tel Aviv, Ramat Aviv 69978, Israel*
e-mail: yaro@eng.tau.ac.il

[2] *Physics Department, Fisk University*
*1000 17th St., Nashville, TN 37298, USA*
e-mail: hjc@fisk.edu

## Abstract

Optics-less cutaneous ("skin") vision is not rare among living organisms, though its mechanisms and capabilities have not been thoroughly investigated. This paper demonstrates, using methods from statistical parameter estimation theory and numerical simulations, that an array of bare sensors with a natural cosine-law angular sensitivity arranged on a flat or curved surface has the ability to perform imaging tasks without any optics at all. The working principle of this type of optics-less sensor and the model developed here for determining sensor performance may be used to shed light upon possible mechanisms and capabilities of cutaneous vision in nature.

# 1  Introduction

Organisms in nature use many types of imaging or visual systems (Parker, 2003) and (Land & Nilsson, 2001). Most of them use optics to form images, but optics-less cutaneous vision (skin vision) is also found among many types of living organisms. One example may be heliotropism, the ability of some plants to orient their leaves or flowers towards the sun. It seems obvious that this ability requires some sort of "skin" vision to determine direction to the sun and direct their flowers or leaves accordingly. There are also many examples of animals with organs that may be classified as optics-less imaging sensors, which do not use optics that focus light. These include:

- "Primitive eyes" of many invertebrates, including eye spots (patches of photosensitive cells located on the skin), cup eyes, and pit eyes (Parker, 2003) and (Land & Nilsson, 2001).

- Cutaneous photoreception in reptiles (Schwab, 2005) and (Zimmerman & Heatwole, 1990).

- "Pit organs" of vipers located near their normal eyes – these organs are sensitive to infra-red radiation and do not contain any optics (Molenaar, 1992) and (Sichert, Friedel, & van Hemmen, 2006).

- The "lateral line system", an organ characteristic of fish, which is sensitive to pressure waves in water and which they use to localize sources of vibration located within approximately one body length (Schellart & Wubbels, 1998).

- Electro-receptors, found in some types of fish and sharks, as well as the duckbilled platypus, are sensitive to electric fields. These organs allow animals to sense electrical variations in their surroundings within approximately one body length. Data presented in (Pettigrew, 1999) suggest that the platypus as a whole has approximately 100 times the sensitivity of a single electroreceptor to an electrical source, which suggests that higher-level processing is being carried out by the platypus.

There are also a number of reports on the phenomenon of cutaneous vision in humans (Gardner, 1966), (Bongard & Smirnov, 1965), and (Zavala, Van Cott, Orr, & Smal, 1967). In particular, (Bongard & Smirnov, 1965) provides quantitative data on the ability of a certain young woman to "see" images using only the fingers of her right hand. It reports also that the subject demonstrated an ability to detect colors, to resolve patterns in near-contact with her fingers with a resolution of about 0.6 mm and to determine simple patterns within a maximum distance of 1-2 cm from the fingers. Refs. (Bach-y-Rita, Tyler, & Kaczmarek, 2003) and (Sampaio, Maris, & Bach-y-Rita, 2001) report experiments suggesting that photosensitive cells on the skin (abdomen, back, foot, and tongue have all been used) may be used to allow blind people to "see" to a limited extent in a process that is termed "sensory substitution." However, most of the available publications provide phenomenological findings rather then focus on mechanisms and capabilities of cutaneous vision.

The properties of cutaneous vision reported in the literature are in a good correlation with properties of "Optics-less Smart" (OLS) sensors proposed recently in (Caulfield, Yaroslavsky, Ludman, 2007), (Caulfield, Yaroslavsky, 2007), and (Caulfield, Yaroslavsky, Goerzen, & Umansky, 2008), which enables us to consider their operational principles as a possible mechanism of cutaneous vision, as well as of other types of primitive vision.

## 2 Method

### 2.1 OLS-sensors: and their operation principle

OLS-sensors consist of arrays of sub-sensors arranged without any optics on a flat or curved surface and supplemented with a signal processing unit. This unit collects outputs of the sub-sensors and computes from them statistically optimal estimates of radiation source intensities and coordinates. The sub-sensors' angular sensitivity is defined by their surface properties. We will assume here the Lambertian model of cosine-law angular sensitivity function since it serves as a good example of many surface types. Depending on a priori knowledge about radiation sources these flat OLS sensors can be used in the following three basic operation modes:

- **"General localization" mode**: estimation of intensities and locations of a known number of point radiation sources

- **"Constellation" mode**: estimation of intensities and locations of "constellations" of radiation sources that consist of a known number and configuration of point sources with known relative distribution of intensities

- **"Imaging" mode**: estimation of intensities of a given number of point radiation sources with known spatial locations

We will explain the operational principle of flat OLS-sensors using the 2-D model shown in Fig. 1.

*** FIGURE 1 ***

Let $\{I, x_0, y_0\}$ be intensity and coordinates of a radiation source in the sensor's coordinate system and let $N$ elementary sub-sensors arranged over a plane (Fig. 1) be used to determine these parameters. Assume that each elementary sub-sensor has angular sensitivity that is proportional to the cosine of the angle between the normal surface and the direction to the source and that its response to the radiation is corrupted by a random component that can be modeled as additive signal independent normal noise. Assume also that the signal processing unit of the sensor is programmed to compute the Maximum Likelihood (ML) estimates of the parameters $\{\hat{I}, \hat{x}_0, \hat{y}_0\}$ (Van-Trees, 1968). For the selected model, ML estimates are solutions to the equation:

$$\{\hat{I}, \hat{x}_0, \hat{y}_0\} = \arg\min_{\{I, x_0, y_0\}} \left\{ \sum_{k=1}^{N} \left[ s_k - I \frac{\cos(\theta_k(x_0, y_0))}{\sqrt{(x_k - x_0)^2 + y_0^2}} \right]^2 \right\}, \qquad (2.1)$$

where:

$$s_k = I \frac{\cos(\theta_k(x_0, y_0))}{\sqrt{x_0^2 + y_0^2}} + n_k, \text{ and } \cos(\theta_k(x_0, y_0)) = \frac{y_0}{\sqrt{(x_k - x_0)^2 + y_0^2}} \qquad (2.2)$$

and $x_k$ - is the *k*-th sub-sensor *x*-coordinate and $n_k$ - random variable with normal distribution that represents sensor's noise.

## 2.2 Theoretical analysis:

The performance of the OLS sensors can be evaluated in terms of standard deviation of the estimation errors. Theoretical lower bounds of the estimation error variance are Cramer-Rao lower bounds (Van-Trees, 1968). Theoretical analysis carried out for the simple case of three sub-sensors and one source at Cartesian coordinates $(x_0, y_0)$ (in units of inter-sub-sensor distance, see Fig. 1, b), gives the following Cramer-Rao Lower Bounds (CRLB) for error variances in estimating source intensity

$$Var\{\hat{I}_{err}\} = Var\{|I - \hat{I}|\} \geq 2\sigma^2 \qquad (2.3)$$

$$Var\{\hat{x}_{err}\} = Var\{|x_0 - \hat{x}_0|\} \geq \frac{2\sigma^2}{I^2} = \frac{2}{SNR} \qquad (2.4)$$

$$Var\{\hat{y}_{err}\} = Var\{|y_0 - \hat{y}_0|\} \geq \frac{\sigma^2}{I^2} = \frac{1}{SNR} \qquad (2.5)$$

where $\sigma^2$ is variance of sub-sensors' noise (in units of source intensity dynamic range) and SNR is defined to be $I^2/\sigma^2$.

## 2.3 Numerical simulation of OLS sensors

We studied how the estimation error variances depend on the model parameters (the number of sub-sensors, the number of sources, distance from sensor to sources, sub-sensor noise variance) by means of statistical simulation, using the multi-start algorithm based on Quasi-Newton optimization procedure in order to solve the optimization problem of Eq. 2.1. For each parameter vector, the model was run 50 times in to accumulate statistics of the performance in terms of standard deviation of the estimation error of the runs. In order to eliminate the results of non-converged runs, a robust estimator of the mean $\tilde{\mu}$ and a robust estimator of the standard deviation $\tilde{\sigma}$ were used:

$$\tilde{\mu} = Q2; \quad \tilde{\sigma} = Q3 - Q1, \tag{2.6}$$

where **Q1**, **Q2** and **Q3** are first, second and third quartiles of the distribution. The points falling out of the range: $[\tilde{\mu} - 4\tilde{\sigma}, \tilde{\mu} + 4\tilde{\sigma}]$ are removed, and then the mean and standard deviation are re-calculated from the remaining points.

## 3  Results

Figs. 2 - 3 present some simulation results. Fig. 2 illustrates the work of the sensor in "imaging" mode and shows results of reconstruction of the intensity of sources arranged in a flat array of 8x16 sources as it is estimated by the flat sensor with 8x16 sub-sensors at different distances from the sources. Noise standard deviation was set to 0.01 (SNR=100). The source intensities form an image containing the characters "SV."

\*\*\* FIGURE 2 \*\*\*

One can vividly see from the figure that the flat OLS-sensor with a realistic signal-to-noise ratio of 100 is capable of reasonably good imaging for sources situated in its close proximity, though the quality of the estimates rapidly decays with its distance to sources. This phenomenon matches observations published in the above-mentioned studies of cutaneous vision.

Fig. 3 enables quantitative evaluation of the size and shape of the sensor's field of view in terms of the standard deviation of estimates of X-Y coordinates and intensity of a single source situated in different positions. In particular, one can see that within small distances from the sensor, accurate estimation of source coordinates and intensity is possible, but at distances greater then about half of the sensor length, the estimation accuracy drops by an order of magnitude. Experiments also revealed that the sensor's capability to resolve close sources is limited by the value of the order of inter-sub-sensor distance, which intuitively is well understood. Experiments with different numbers of sub-sensors show similar results. Comparing the flat sensor performance with curved sensor performance shows that the curved sensor outperforms the flat sensor in x- and y-axis estimation, as well as increasing the field of view of intensity estimation. However, the flat sensor has the best performance in estimating intensity in the region directly above the sensor.

*** FIGURE 3 ***

These results demonstrate that modifying the sensor design by bending the flat surface of sub-sensors results in an increased field of view. In (Caulfield, Yaroslavsky, Goerzen, & Umansky, 2008), we show that an array of sub-sensors placed on outer surface of a sphere and supplemented with a corresponding signal processor has an unlimited field of view.

## 4  Discussion and conclusion

The simulation results for OLS sensors presented here show that good directional vision without optics is possible even using the simplest possible sensors whose angular sensitivity is defined by the natural Lambertian surface absorptivity. Main advantages of the OLS-sensors are:

- No optics are needed, making this type of sensor applicable to virtually any type of radiation and to any wavelength.

- The design is flexible: the shape of the sensor's surface can be easily changed to fit the required sensor's field of view without any changes in the signal processing software except updating its input parameters (angles of sub-sensor normals in the sensor coordinate system).

- The sensor's resolving power is determined ultimately by the number of sub-sensors and their signal-to-noise ratio, and not by diffraction-related limits.

The cost for these advantages is the high computational complexity, especially when good imaging properties for multiple sources are required. This might be a reason why natural selection resulted in lens-based camera vision rather than "skin" vision as a main vision instrument when high resolution within sufficiently large field of view is required. What a lens does in parallel and with a speed of light, optics-less vision must replace by computations in neural machinery, which is slower and requires high energy (food) consumption. Although evolutionary biology is beyond the scope of this paper, we take the liberty to suggest that cutaneous vision may have appeared in very early stages of evolution and diverged into two forms. Through bending of the sensor's surface to convex spherical form and by sharpening the angular sensitivity of elementary sub-sensors by means of lenslets and waveguides, the eyespot may evolved into the compound facet eye, in which the complexity of the required neural machinery is proportional to the number of facets. Through bending of the sensor's surface to concave spherical form, the eye spot and its neural machinery may evolved to camera-like vision coupled with a brain. Thanks to the processing achieved by the lens, the brain does not need to spend much of its resources on image formation and instead can focus on image understanding.

To conclude, these results show that good directional vision is possible even using the simplest possible sensors whose angular sensitivity is defined by the natural Lambertian surface absorptivity. We believe that the operational principle and properties of OLS sensors may cast a new light on mechanisms, advantages and limitations of extra-ocular vision in nature and may suggest new ways for planning further experiments with animals and humans.

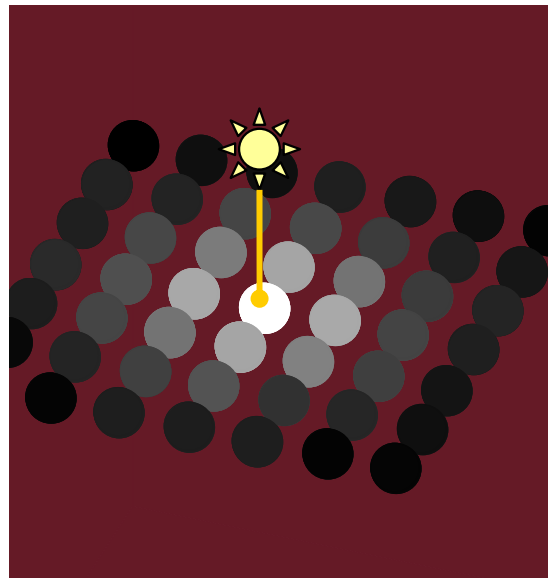

a)

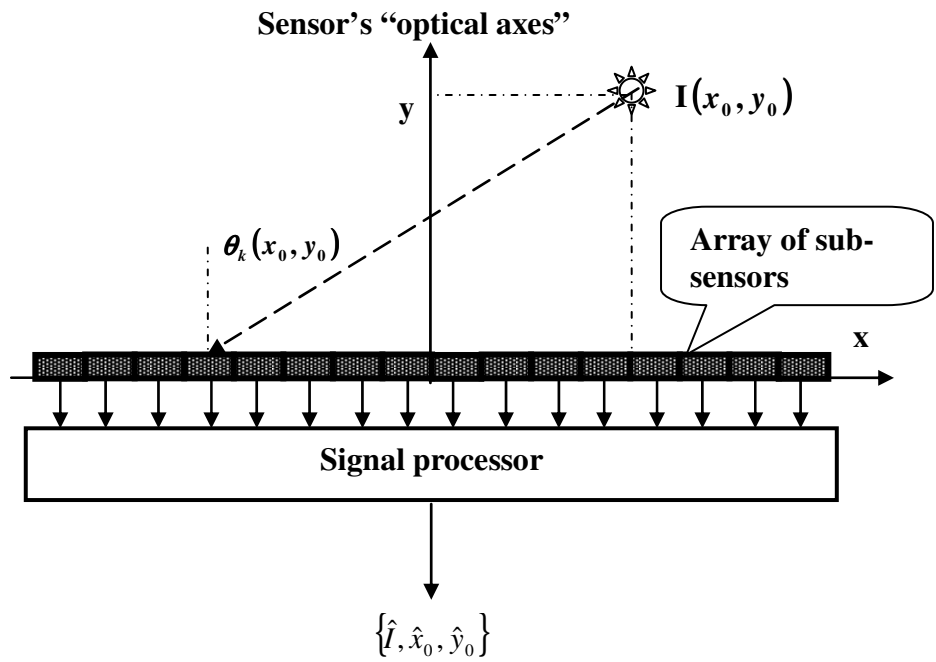

b)

Fig. 1: Flat OLS sensor (a) and its schematic diagram (b)

**Figure 2**

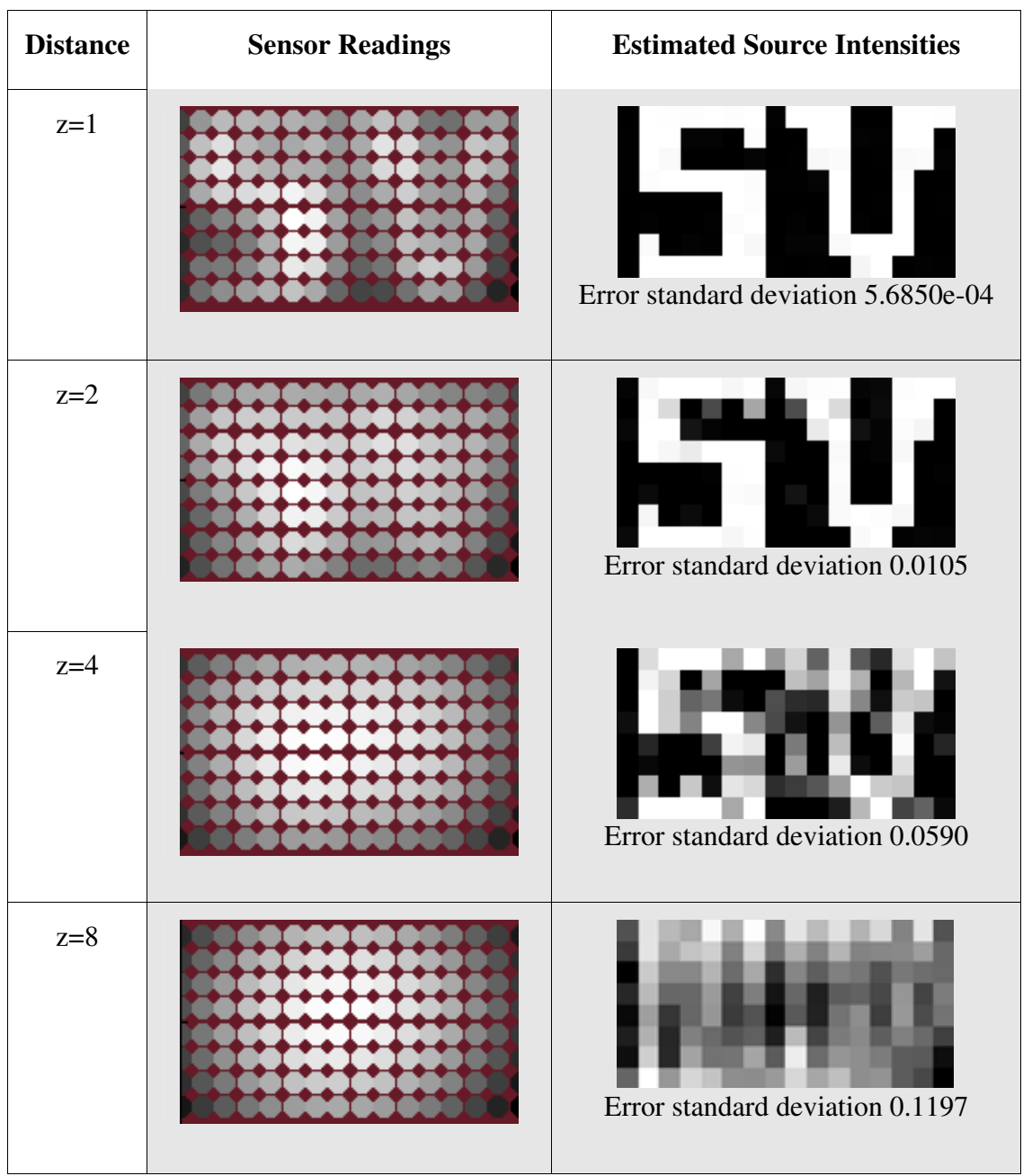

Fig. 2: An illustration of sensing of 8x16 radiation sources arranged on a plane in form of characters "SV" by a 3-D model of a flat OLS sensor of 8x16 elementary sub-sensors in the "imaging" mode for distances of sources from the sensor $Y$=1 to 8 (in units of inter-sub-sensor distance). SNR was kept constant at 100 by making the source amplitude proportional to the distance between the source plane and sensor plane. Sub-sensor noise was 0.01.

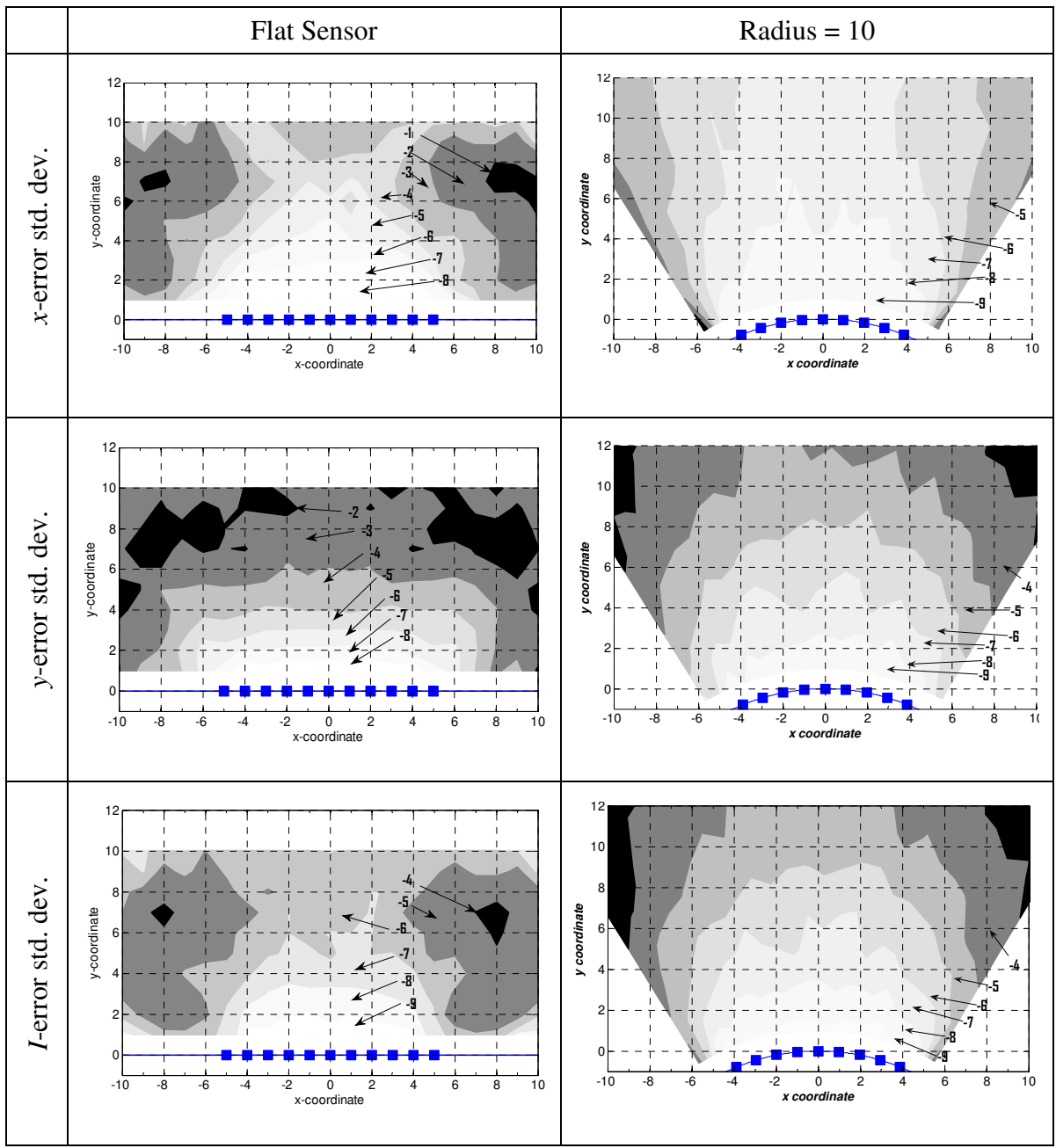

Fig. 3. Sensors on flat (left) and bent (right) surfaces: Standard deviations of estimation errors of intensity, *I,* and of X-Y coordinates, *x* and *y*, of a single source as a function of the source position with respect to the sensor. Numerals marked on the contours indicate the exponent of 2 of the numerical values of the error standard deviation (in same units).